\documentclass[aps,prb,twocolumn,showpacs,floats]{revtex4}
\usepackage{epsfig}
\usepackage{amsmath}
\newcommand{\be}{\begin{equation}}
\newcommand{\ee}{\end{equation}}
\newcommand{\bea}{\begin{eqnarray}}
\newcommand{\eea}{\end{eqnarray}}
\newcommand{\ba}{\begin{array}}
\newcommand{\ea}{\end{array}}
\newcommand{\nn}{\nonumber}
\newcommand{\parti}{\partial}
\newcommand{\Del}{\Delta}
\newcommand{\del}{\delta}
\newcommand{\Gam}{\Gamma}
\newcommand{\gam}{\gamma}
\newcommand{\al}{\alpha}

\newcommand{\eps}{\epsilon}

\begin{document}

\title{\bf Charge Pumping in Mesoscopic Systems coupled to a Superconducting Lead} 

\author{M. Blaauboer
} 

\affiliation{ Department of Applied Physics, Delft University of Technology, 
Lorentzweg 1, 2628 CJ Delft, The Netherlands}
\date{\today}

\begin{abstract}
We derive a general scattering-matrix formula for the pumped current through a
mesoscopic region attached to a normal and a superconducting lead. As applications of this
result we calculate the current pumped through (i) a pump in a wire, (ii) a
quantum dot in the Coulomb blockade regime, and (iii) a ballistic double-barrier 
junction, all coupled to a superconducting lead. Andreev reflection is shown
to enhance the pumped current by up to a factor of 4 in case of equal coupling
to the leads. We find that this enhancement can still be further
increased for slightly asymmetric coupling.
\end{abstract}

\maketitle

\section{Introduction}
\label{sec-intro}
It is well-known that the quantum transport properties of a mesoscopic system
are modified in the presence of a superconducting interface, due to
interference between normal and Andreev reflections. Andreev
reflection (AR) \cite{andr64} is the electron-to-hole reflection process which occurs
when an electron with energy slightly above the Fermi energy is incident on
the boundary between a normal metal and a superconductor: the electron enters
the superconductor after forming a Cooper pair, and leaves a hole in the
normal metal with energy slightly below the Fermi level which travels back
along (nearly) the same path where the electron came from. Because of the
phase-coherent character of AR, it is interesting to study its effect on
transport in mesoscopic systems, where phase coherence plays an important
role. In the last decade, this has led to 
the discovery of a wealth of quantum
interference effects in mesoscopic normal-metal--superconductor (NS)
structures \cite{been97}, such as the observation of a 
large narrow peak in the differential 
conductance of a disordered NS junction 
("reflectionless tunneling") \cite{kast91}, and the discovery 
of novel Kondo phenomena in quantum dots attached to a normal and a 
superconducting lead \cite{kang98}. In addition, 
investigations of the conductance in 
superconductor--carbon-nanotube devices have recently appeared\cite{morp99}, 
which indicate that also in these devices resonant behavior due to AR 
occurs. \newline 
\indent  The purpose of this paper is to study the effects of AR on a new type
of mesoscopic transport, namely adiabatic quantum pumping. Quantum pumping involves the
generation of a d.c. current in the absence of a bias voltage by periodic
modulations of two or more system parameters, such as e.g. the shape of the system or a
magnetic field. The idea was pioneered by Thouless for electrons moving in an
infinite one-dimensional periodic potential \cite{thou83}. 
In recent years, adiabatic quantum pumping in 
quantum dots has attracted a lot of 
attention \cite{brou98,alei98,swit99}. Quantum dots 
are small metallic or semiconducting
islands, confined by gates and connected to 
electron reservoirs (leads) through quantum point contacts (QPCs) \cite{kouw97}. 
In addition to investigations of pumping in quantum dots, 
theoretical ideas have been put forward for charge pumping in 
carbon nanotubes \cite{wei01}, and for pumping
of Cooper pairs \cite{zhou99}. Here we consider a mesoscopic system consisting 
of an arbitrary normal-metal 
region, e.g. a quantum dot or QPC, coupled to a superconductor, as schematically
depicted in Fig.~\ref{fig:system}(a). We start by
deriving a general formula for the pumped current through this NS system in terms 
of its scattering matrix (Sec.~\ref{sec-deriv}). This is the $N$-mode 
generalization of the result of 
Wang {\it et al.} for a NS system with single-mode leads\cite{wang01}. 
We then use this result to calculate the 
current pumped through: a simple peristaltic pump (Sec.~\ref{subsec-wire}), 
a quantum dot in the 
Coulomb blockade regime (Sec.~\ref{subsec-dot}), and a
double-barrier junction (Sec.~\ref{subsec-ballistic}), each coupled to a
superconducting lead. Comparing these with the pumped current in the
corresponding systems attached to normal leads only, shows that AR enhances 
quantum pumping by up to a factor of $>4$ in systems with (nearly) 
symmetric coupling to the leads, while it reduces 
quantum pumping in the opposite situation of strongly asymmetric coupling. 

\section{Derivation of NS pumping formula}
\label{sec-deriv}

\begin{figure}
\centerline{\epsfig{figure=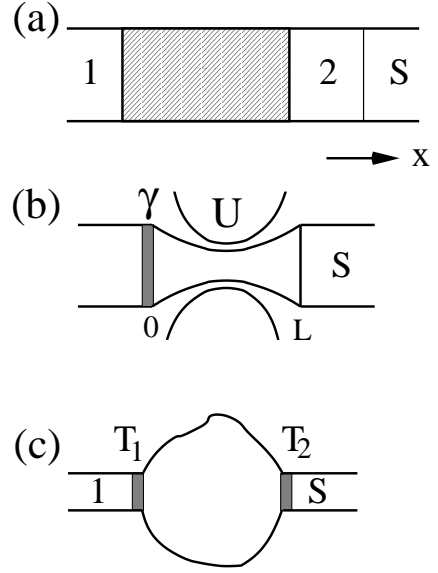,width=0.65\hsize}}
\caption[]{(a) Normal-metal region (hatched) adjacent to a superconducting lead (S). The
  normal leads 1 and 2 contain N modes each. (b) NS pump in a 1D wire containing
  a tunnel barrier $\gam$ and an external potential $U$. (c) Quantum dot coupled via
  tunneling barriers $T_1$ and $T_2$ to a normal (1) and a superconducting (S)
  lead.
}
\label{fig:system}
\end{figure}

Consider the system in Fig.~\ref{fig:system}(a). The normal region 
(which may e.g. be disordered, or contain a
constriction) is coupled to ideal normal leads
1 and 2 containing $N$ modes each. No bias voltage is applied to
the system, so all reservoirs are held at the same potential. 
We assume a constant pair potential $\Del(\vec{r})= \Del_0 e^{i\phi}$ 
in the superconductor, which is 
applicable for wide junctions\cite{likh79} and has 
previously been used to derive the conductance through a NS
junction\cite{been92}. We also assume that the NS interface is ideal, 
i.e. no specular reflection occurs for energies $0<\eps < \Del_0$, 
with $\eps$ the energy measured from the Fermi energy $\eps_F$. 
The scattering matrix $S_{\rm NS}$ of the entire system is given
by\cite{been94}
\bea
S_{\rm NS}(\eps) = \left( \ba{ll}
S^{ee}(\eps) & S^{eh}(\eps) \\
S^{he}(\eps) & S^{hh}(\eps)
\ea \right),
\label{eq:Smatrix}
\eea
where $S^{ee}$-$S^{hh}$ are $N\times N$ scattering matrices given by
\begin{subequations}
\bea
S^{ee}(\eps)& =& r_{11}(\eps) +
\al^2 t_{12}(\eps)r_{22}^{*}(-\eps)M_{e}t_{21}(\eps), \\
S^{eh}(\eps)&=& \al e^{i\phi} t_{12}(\eps) M_{h} t_{21}^{*}(-\eps), \\
S^{he}(\eps)&=& \al e^{-i\phi} t_{12}^{*}(-\eps) M_{e} t_{21}(\eps), \\
S^{hh}(\eps)& =& r_{11}^{*}(-\eps) +
\al^2 t_{12}^{*}(-\eps) r_{22}(\eps) M_{h} t_{21}^{*}(-\eps).  
\eea
\label{eq:Selements}
\end{subequations}
Here $r_{ii}(\eps)$ [$t_{ij}(\eps)$], $i,j=1,2$, denotes the reflection
[transmission] amplitude for electrons at energy $\eps$, with $0 < \eps <
\Del_0$, traveling from lead $i$ [$j$] to lead $i$, $\al \equiv \exp[-i \arccos
(\eps/\Del_0)]$, $M_e \equiv [1 - \al^2 r_{22}(\eps) r_{22}^{*}(-\eps)]^{-1}$, 
and $M_h \equiv [1 - \al^2 r_{22}^{*}(-\eps) r_{22}(\eps)]^{-1}$. The scattering
matrix (\ref{eq:Smatrix}) is unitary and satisfies the
symmetry relation $S_{\rm NS}(\eps,B,\phi)_{ij}=S_{\rm NS}(\eps,-B,-\phi)_{ji}$ for
time reversal invariance. Adiabatic
quantum pumping in this NS junction is obtained by slow and periodic
variations of two external parameters $X_1$ and $X_2$ as $X_1(t) = \bar{X}_1 +
\del X_1 \sin(\omega t)$ and $X_2(t) = \bar{X}_2 + \del X_2 \sin(\omega t + \phi)$. 
The frequency $\omega$ has to be such that $\omega \ll \tau_{\rm dwell}^{-1}$, 
with $\tau_{\rm dwell}$ the time particles spend in the system, in order for 
equilibrium to be maintained throughout the entire
pumping cycle. The net charge $\del Q(t)$ emitted into lead 1 due to the 
modulations $\del X_1$ and $\del X_2$ consists of the amount of negative charge
carriers (electrons) minus the amount of positive charge carriers (holes)
emitted into lead 1. For infinitesimal variations $\del X_i$, $i=1,2$, 
this charge is given by
\bea
\del Q(t) & = & \frac{e}{2\pi} \sum_{\al, \beta \in 1} \left[ {\rm Im}
\left( \frac{\parti S_{\al \beta}^{ee}}{\parti X_1} S_{\al \beta}^{ee\, *} -
\frac{\parti S_{\al \beta}^{he}}{\parti X_1} S_{\al \beta}^{he\, *} \right)
\del X_1(t) \right.  \nn \\
& & + \left. {\rm Im}
\left( \frac{\parti S_{\al \beta}^{ee}}{\parti X_2} S_{\al \beta}^{ee\, *} -
\frac{\parti S_{\al \beta}^{he}}{\parti X_2} S_{\al \beta}^{he\, *} \right)
\del X_2(t) \right],
\label{eq:chargetime}
\eea
where the indices $\al$ and $\beta$ are summed over all $N$ modes in lead 1.
This expression is obtained along the same lines as the pumped current in a 
quantum dot coupled to two normal leads\cite{brou98} and based on a formula 
derived by B\"uttiker {\it et al.}\cite{buet94}, see also Ref.~\cite{wang01}. 
The total charge emitted into
lead 1 during one period $\tau\equiv 2\pi/\omega$ is found by integrating 
Eq.~(\ref{eq:chargetime}) over time, 
\bea
Q(\tau) & = & \frac{e}{2\pi} \int_0^\tau dt \sum_{\al, \beta \in 1} \left[ 
{\rm Im} \left( \frac{\parti S_{\al \beta}^{ee}}{\parti X_1} S_{\al \beta}^{ee\, *} -
\frac{\parti S_{\al \beta}^{he}}{\parti X_1} S_{\al \beta}^{he\, *} \right)
\right.  \nn \\
& & \frac{d X_1}{dt} + \left. {\rm Im}
\left( \frac{\parti S_{\al \beta}^{ee}}{\parti X_2} S_{\al \beta}^{ee\, *} -
\frac{\parti S_{\al \beta}^{he}}{\parti X_2} S_{\al \beta}^{he\, *} \right)
\frac{d X_2}{dt} \right], \nn
\eea
and rewriting this as an integral over the area $A$ that is enclosed in 
parameter space ($X_1$,$X_2$) during one period. We then find that 
the total current $I_{\rm NS} \equiv \frac{\omega}{2\pi} Q(\tau)$ 
pumped into lead 1 is given by
\begin{subequations}
\bea
I_{\rm NS} &=& \frac{\omega e}{2\pi^2} \int_{A} dX_1 dX_2 \sum_{\al, \beta \in
  1} \Pi_{\al \beta}(X_1,X_2)
\label{eq:NScurr1}\\
& \approx & 
\frac{\omega e}{2\pi} \del X_1 \del X_2 \sin \phi \sum_{\al, \beta \in
  1} \Pi_{\al \beta}(X_1,X_2),
\label{eq:NScurr2}
\eea
\label{eq:NScurr}
\end{subequations}
with
\bea
\Pi_{\al \beta}(X_1,X_2) & \equiv & {\rm Im} \left[ \frac{\parti S_{\al 
\beta}^{ee\, *}}{\parti X_1} \frac{\parti S_{\al \beta}^{ee}}{\parti X_2} - 
\frac{\parti S_{\al
      \beta}^{he\,*}}{\parti X_1} \frac{\parti S_{\al \beta}^{he}}{\parti X_2}
\right].
\eea 
Eq.~(\ref{eq:NScurr}) is valid at
zero temperature and to first order in the frequency $\omega$; For $N$$=$$1$ it reduces
to the single-mode result of Ref.~\cite{wang01} (apart from a factor of 
2 for spin degeneracy in the latter). Eq.~(\ref{eq:NScurr2}) applies for 
bilinear response in the parameters $X_1$ and
$X_2$, in which case the integral in (\ref{eq:NScurr1}) becomes independent of
the pumping contour. $I_{\rm NS}$ is of similar generality as the expression for
the pumped current in the presence of two normal-metal leads\cite{brou98}
\bea
I_{\rm N} = \frac{\omega e}{2\pi^2} \int_A dX_1 dX_2 \sum_{\al \in 1} \sum_{\beta \in
  \{1,2\}} {\rm Im} \left( \frac{\parti S_{\al \beta}^{*}}{\parti X_1}
  \frac{\parti S_{\al \beta}}{\parti X_2} \right).
\eea
Here $S_{\al \beta}$ denotes the $2N\times 2N$ scattering matrix of the 
system. Note that the index $\beta$ in this case is summed over the 
modes in both lead 1 and lead 2, since electrons can be incident from either lead. 
In our NS junction, the charge pumped into the right lead is converted into a 
supercurrent. 

In order to illustrate the result (\ref{eq:NScurr}), we now proceed to
apply it to several NS configurations. Unless otherwise noted, we restrict 
ourselves to the linear response regime corresponding to weak pumping. 
In that regime, only the scattering matrix at the Fermi level $\eps=0$
is needed. 

\section{Applications}

\subsection{Peristaltic pump in a one-dimensional (1D) wire}
\label{subsec-wire}
 
As a first example, consider a "peristaltic" pump formed by a 1D wire in contact 
with a superconducting lead, see Fig.~\ref{fig:system}(b). The pump is operated by
  periodically opening and closing a tunnel barrier of height $\gam$ and 
  varying an external potential $U$. Calculating the
  scattering matrix (\ref{eq:Smatrix}) from the Schr\"{o}dinger equation with
  potential $V(x)= \gam \delta(x) + U \Theta[x(L-x)]$ and substituting into
  Eq.~(\ref{eq:NScurr2}) yields the current
\bea
I_{\rm NS} = \frac{3\sqrt{2}}{32} \frac{\omega e}{\pi k^2} \del U
  \sin^2(kL).
\label{eq:peripump}
\eea
Comparing this with the analogous expression in the case of two normal
  leads\cite{brou98}, 
\bea
I_{\rm N}= \frac{\omega e L}{8\pi^2k} \del U + 
\frac{\omega e}{16 \pi^2 k^2} \del U (\pi \sin^2(kL) - \sin(2kL)),
\eea 
we see that the presence of AR can both enhance and reduce the pumped 
current. Maximum enhancement of $I_{\rm NS}/I_{\rm N} = 3\sqrt{2}/2 
\approx 2.1$ occurs for short wires such that $2kL \ll 1$. Note that 
$I_{\rm NS}$, unlike $I_{\rm N}$, has no classical contribution and is 
entirely due to quantum interference in the wire.

\subsection{ Pumped current peak heights in a nearly-isolated quantum dot}
\label{subsec-dot} 

As a second example, we study quantum pumping in a quantum dot which is coupled 
via two tunneling barriers with transmission probabilities $T_1$ and $T_2$ to 
a normal and a superconducting single-mode lead, see Fig.~\ref{fig:system}(c). 
Pumping is achieved by periodic variations of the strength of the two tunneling 
barriers $V_1$ and $V_2$ (with $V_{1,2} \equiv \sqrt{(1-T_{1,2})/T_{1,2}}$ for
delta-function barriers) as 
$V_1(t) = \bar{V}_1 + \del V_1 \sin(\omega t)$ and $V_2(t) = \bar{V}_2 + \del V_2 
\sin(\omega t + \phi)$. We are interested in the regime of high barriers (where 
transmission is low, $T_1,T_2 \ll 1$), and weak pumping 
($\del V_m \ll \bar{V}_m, m=1,2$). At low temperatures such that 
$k_B T \ll \Del < E_C$ [with $\Delta$ the single-particle level 
spacing and $E_C$$=$$e^2/C$ the charging energy of the dot, $C$ being 
the total capacitance] the quantum dot then remains in the Coulomb 
blockade regime during the whole 
pumping cycle and transport through the dot is mediated by resonant transmission 
through a single level\cite{been91}. Substituting 
the appropriate scattering matrix\cite{blaa01,levi00} into 
Eq.~(\ref{eq:NScurr2}) yields, up to lowest order in $T_1$ and $T_2$ 
and for thermal energies less than the total decay width $\Gam$ into the leads
$k_B T < \Gam \ll \Del$ [with $\Gam = \Gam_1 
+ \Gam_2 \equiv \hbar \nu (T_1 + T_2)$ and $\nu$ the attempt frequency, 
the inverse of the round-trip travel time between the two barriers], 
\bea
I_{\rm NS} = \frac{\omega e}{4 \pi} \del V_1 \del V_2 \sin \phi \frac{T_1^{7/2} 
T_2^{5/2} (\sqrt{T_1} + \sqrt{T_2})}{[\frac{1}{4}(T_1^2 + T_2^2) +  \left( 
\frac{\eps - \eps_{\rm res}}{\hbar \nu} \right)^2]^3}.
\label{eq:peakcurr}
\eea
Here $\eps_{\rm res}$ denotes the resonance energy for a completely isolated dot 
($T_1$$=$$T_2$$=$$0$). Note that Eq.~(\ref{eq:peakcurr}) is not symmetric 
with respect to $T_1$ and $T_2$, in contrast 
with the conductance\cite{been92} $G_{\rm NS}$$=$$\frac{e^2}{h}\, 
\frac{T_1^2 T_2^2}{[\frac{1}{4} (T_1^2 + T_2^2) + \left( 
\frac{\eps - \eps_{\rm res}}{\hbar \nu} \right)^2]^2}$ 
through this system. This 
is due to the fact that $I_{\rm NS}$ depends on $\parti S / \parti X$, 
whereas $G_{\rm NS}$ depends on the transmission eigenvalues, the eigenvalues
of the matrix $t_{12}^{\dagger}t_{12}$.
Compared to the pumped current in a dot coupled to two normal leads\cite{blaa01}
\bea
I_{\rm N} = \frac{\omega e}{4 \pi} \del V_1 \del V_2 \sin \phi \frac{(T_1 T_2)^{3/2} 
(\sqrt{T_1} + \sqrt{T_2}) (T_1 + T_2)}{[\frac{1}{4}(T_1 + T_2)^2 +  \left( 
\frac{\eps - \eps_{\rm res}}{\hbar \nu} \right)^2]^2},
\eea
we find that
\bea
\frac{I_{\rm NS}}{I_{\rm N}}= \frac{ T_1^2 T_2 \left[ \frac{1}{4}(T_1 + T_2)^2 + \left( \frac{\eps - \eps_{\rm res}}{\hbar \nu} \right)^2 \right]^2 }{ (T_1 + T_2) \left[ \frac{1}{4} (T_1^2 + T_2^2) + \left( \frac{\eps - \eps_{\rm res}}{\hbar \nu} \right)^2 \right]^3 }.
\label{eq:ratio} 
\eea
For symmetric coupling ($T_1$$=$$T_2$) close to resonance, $(\eps - \eps_{\rm res}) \ll \hbar 
\nu T_{1,2} \equiv \Gam_{1,2}$, AR enhances the pumped current by a factor of 
4~\cite{previous}. In case of strongly asymmetric coupling close to resonance,
on the other hand, AR reduces the pumped amplitude: 
$I_{\rm NS}/I_{\rm N} \sim (T_1/T_2)^2$$ \ll $$ 1$ for $T_1$$ \ll $$ T_2$, and 
$I_{\rm NS}/I_{\rm N} \sim T_2/T_1$$ \ll$$ 1$ for $T_2$$ \ll $$T_1$. The enhancement by a 
factor of 4 for symmetric barriers consists of 2 contributions of a factor of 2: one factor of 
2 is due to the contribution of both electrons and holes to the current, which is 
also responsible for the doubling of conductance $G_{\rm NS}/G_{\rm N}$$=$$2$ in this NS 
structure\cite{been92}. A second factor of 2 comes from the 
asymmetry of the NS dot with respect to injection of charge carriers into the leads, since 
electrons can only leave the system through the left, normal lead. This leads to 
an extra doubling of the pumped current compared to the normal case where electrons 
are injected into both the left and
the right leads. This extra factor of 2 does not occur in the presence of an 
applied bias, as for conductance,
since the bias causes charge carriers to flow from one side to the other in both the 
normal and the NS system. Note that due to the asymmetry of Eq.~(\ref{eq:ratio}) 
with respect 
to $T_1$ and $T_2$ the maximum attainable enhancement is even larger than 4: for a slightly 
asymmetric junction (with $T_1/T_2$$ \sim $$ 1.26$) one obtains $I_{\rm NS}$/$I_{\rm N}$$\sim$
4.23. In this case quantum interference between electrons and holes in the NS system is maximal.
If the barrier asymmetry is further increased, $I_{\rm NS}/I_{\rm N}$ decreases and 
eventually becomes less than 1 for strongly asymmetric coupling, when pumping is dominated 
by one barrier only.

At temperatures higher than the decay width, $\Gam$$ \ll$$ k_B T$$ \ll$$ \Del$, the pumped current
exhibits Coulomb oscillations as a function of an applied gate voltage\cite{blaa01}. 
The peak heights of 
these oscillations can be obtained by thermally averaging Eq.~(\ref{eq:peakcurr}) 
as $I_{\rm NS,peak} 
$$\equiv $$- \int d\eps\, I_{\rm NS} f^{'}(\eps,T)$$ \approx $$\frac{1}{4k_{B} T} \int_{\eps_{\rm res}-\frac{1}{2}\hbar \nu \sqrt{T_1^2 + T_2^2}}^{\eps_{\rm res}} d\eps\, I_{\rm NS}$, where $f(\eps,T)$$ \equiv $$
[1 + \exp(\eps/k_B T)]^{-1}$ denotes the Fermi function. We obtain 
\bea
I_{\rm NS,peak} = \frac{\omega e (8+3\pi)\hbar \nu }{16\pi k_B T} \del V_1 
\del V_2 \sin \phi\, \frac{T_1^{\frac{7}{2}} T_2^{\frac{5}{2}} (T_1^{\frac{1}{2}} + T_2^{\frac{1}{2}})}{(T_1^2 + T_2^2)^{\frac{5}{2}}}.
\label{eq:currpeak2}
\eea
This thermal average does not explicitly include the effect of the charging energy $E_C$ 
on the pumped 
current. A full linear response theory for Coulomb blockade conductance oscillations including charging energy 
was developed in Ref.~\cite{been91}. There it was shown that for temperatures $k_{B}T \ll \Del$ 
only one level $N_{\rm min}$ participates in the transport [$N_{\rm min}$ is 
defined as the level which minimizes the energy $E_{\rm N}$+U(N)-U(N-1)-
$\eps_{\rm F}$, with $E_{\rm N}$ the energy of the Nth level of the dot, and U(N)
the electrostatic energy of a dot containing N electrons], and the 
oscillation peaks are well described by the thermal average. 
One can show that for the same reason the pumped current 
peaks in this temperature range 
are well described by the thermal average (\ref{eq:currpeak2}), with 
the understanding that 
$T_{1,2}$ in Eq.~(\ref{eq:currpeak2}) refer to the level $N_{\rm min}$. 
From (\ref{eq:currpeak2}) and the analogous normal-state result\cite{blaa01}
\bea 
I_{\rm N,peak} = \frac{\omega e (2+\pi)\hbar \nu }{16\pi k_B T} \del V_1 
\del V_2 \sin \phi\, \frac{(T_1 T_2)^{\frac{3}{2}} (T_1^{\frac{1}{2}} + 
T_2^{\frac{1}{2}})}{(T_1 + T_2)^2},
\eea
we obtain
\bea
\frac{I_{\rm NS, peak}}{I_{\rm N, peak}} = \left( \frac{8+3\pi}{2+\pi} \right) 
\frac{T_1^2 T_2 (T_1 + T_2)^2}{(T_1^2 + T_2^2)^{5/2}}.
\label{eq:ratio2}
\eea 
Also here, AR enhances the pumped current in case of symmetric tunnel barriers, while 
a reduction occurs for asymmetric barriers. Maximum enhancement of 
$I_{\rm NS, peak}/I_{\rm N, peak}$ $\sim $ 2.55 is reached for 
$T_1/T_2$$ \sim$ 1.292. This factor is less than 4, because the average over energy $\eps$ from which the peak heights (\ref{eq:currpeak2}) are obtained also involves contributions
of $I_{\rm NS}$ [Eq.~(\ref{eq:peakcurr})] further away from resonance, for which $I_{\rm NS}/I_{\rm N}$ is much less than 4 [consider e.g. Eq.~(\ref{eq:ratio}) for $\eps - \eps_{\rm res}
= \frac{1}{2} \hbar \nu \sqrt{T_1^2 + T_2^2}$]. This results in lower maximal enhancement
of the pumped current peaks (\ref{eq:ratio2}) at higher temperatures $k_{B}T \gg \Gam$.

Another interesting result is obtained in this system by relaxing the assumption of weak pumping 
and considering quantum pumping by varying the two tunneling barriers in such a way that the loop which describes the pumping 
cycle in parameter space encircles the entire resonance line. For normal-metal contacts 
this problem has recently been studied\cite{levi00} and led to the prediction that at zero temperature the charge transferred during one pumping cycle is quantized, $Q=e$ (for spinless electrons). The transferred charge in our NS system 
is obtained by substituting the scattering matrix of a 1D double-barrier junction given in Ref.~\cite{blaa01} into Eq.~(\ref{eq:NScurr}a) and integrating over the resonance line 
$V_1^{-1} + V_2^{-1} = |(\eps - \eps_{\rm res})/\hbar \nu| \ll 1$, with $V_i^{-1} \equiv \sqrt{T_i/(1-T_i)}$ for $i=1,2$. We then obtain
\bea 
Q=\frac{3\sqrt{2}e}{2} \int_{-1}^{1} dz \frac{(1+z)^2(1-z^2)^3}{[1+6z^2+z^4]^{5/2}}
 = e,
\eea 
so AR neither enhances nor reduces quantum pumping. This occurs because charge is 
effectively transferred by a shuttle mechanism (first through one barrier and
then through the next), which is unaffected by Andreev interference effects
and fixed by the pumping loop. Since only pairs of electrons can enter the superconductor, but the strong Coulomb interaction (charging energy) forbids simultaneous pumping
of 2 electrons with opposite spin, this pumping process is {\it not allowed} in a nearly-closed NS quantum dot. As pointed out in Ref.~\cite{wang2} it can, however, occur in a double-barrier junction in which electron-electron interactions may be neglected. Both in case of two normal and in case of one normal and one superconducting single-mode lead a quantized
amount of charge of 2e is then transferred during each pumping cycle\cite{wang2}.

\subsection{Ballistic double-barrier junction}
\label{subsec-ballistic} 

\begin{figure}
\centerline{\epsfig{figure=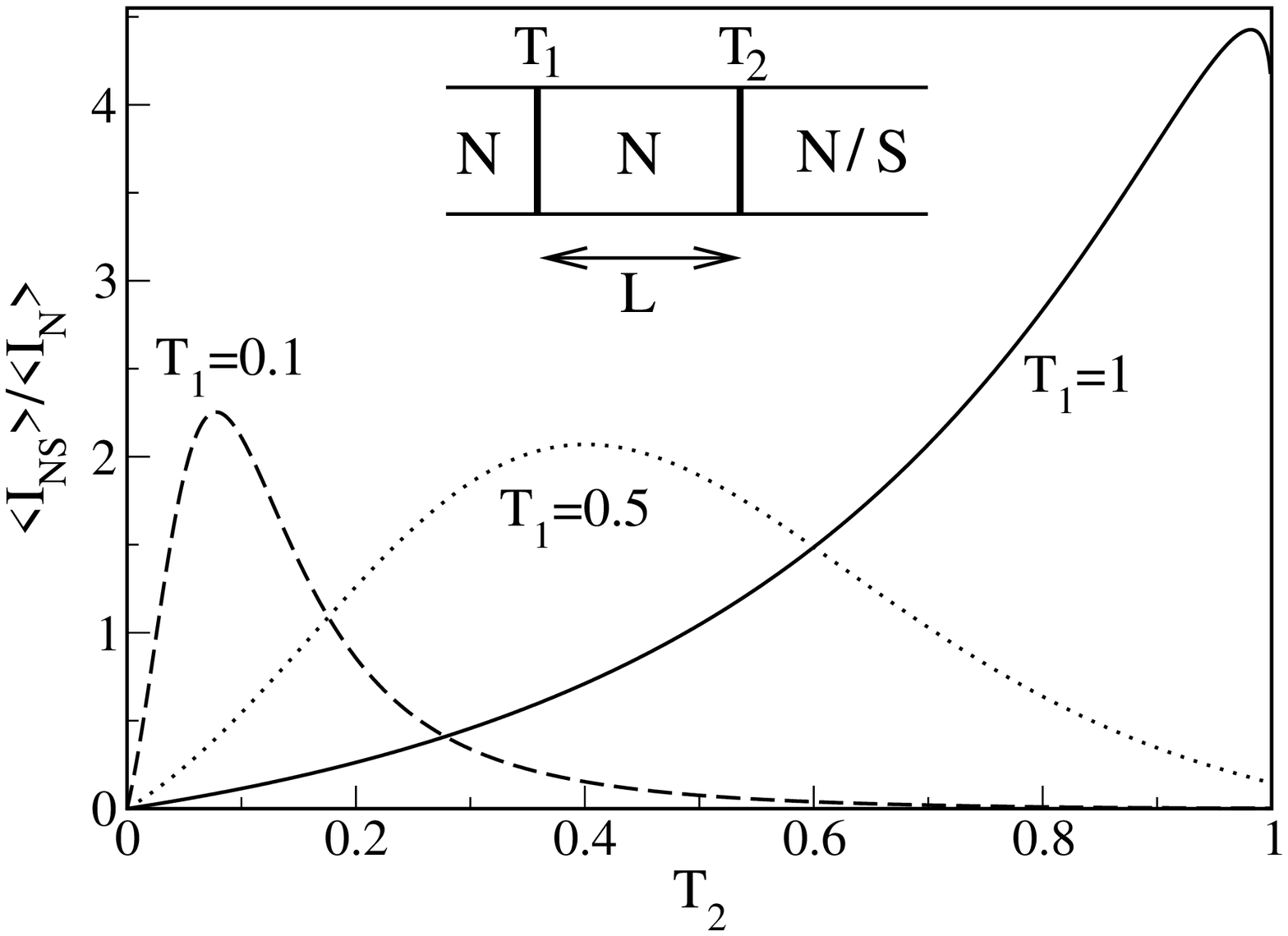,width=0.8\hsize}}
\caption{Ratio of the average pumped currents  $\langle I_{NS}\rangle $ and 
$\langle I_N \rangle$
in ballistic NININ and NINIS junctions as a function of $T_2$ for $T_1$=1,
 0.5, and 0.1. The inset shows the double-barrier junction considered.  
}
\label{fig:doublebarrier}
\end{figure}

Finally, we compare the pumped current $I_{\rm N}$ in a $N I_1 N I_2 N$ junction vs. $I_{\rm NS}$ in a $N I_1 N I_2 S$ junction, where $I_{1,2}$ denote tunnel barriers with transmission probability per mode $T_{1,2}$, see inset of Fig.~\ref{fig:doublebarrier}. In linear response, assuming $N$-mode leads and ballistic transport between the barriers, the pumped currents are given by
\bea
I_{\rm N} =  I_{\rm C} T_1^{\frac{3}{2}} T_2^{\frac{3}{2}} 
\sum_{n=1}^{N} \frac{A_{\rm N} + B_{\rm N} \cos \phi_n + C_{\rm N} \sin \phi_n}{(D_{\rm N}+
E_{\rm N} \cos \phi_n)^2},
\label{eq:double1}
\eea
and
\bea
I_{\rm NS} = 4 I_{\rm C} T_1^3 T_2^{\frac{5}{2}} 
\sum_{n=1}^{N} \frac{A_{\rm NS} + B_{\rm NS} \cos \phi_n + C_{\rm NS} \sin \phi_n}
{(D_{\rm NS}+
E_{\rm NS} \cos \phi_n)^3}.
\label{eq:double2}
\eea
Here $I_{\rm C} \equiv \frac{\omega e}{2\pi} \del V_1 \del V_2 \sin \phi$\cite{spindegen}, 
and 
$A_{\rm N},\dots,E_{\rm NS}$ are given by $A_{\rm N}\equiv\sqrt{R_1 T_2}+\sqrt{R_2 T_1}$, 
$B_{\rm N}\equiv-\sqrt{R_1 R_2}(\sqrt{R_1 T_2} + \sqrt{R_2 T_1})$, 
$C_{\rm N}\equiv -1 + R_1 R_2 - \sqrt{R_1 R_2 T_1 T_2}$, $D_{\rm N}\equiv 1 + 
R_1 R_2$, $E_{\rm N}\equiv -2 \sqrt{R_1 R_2}$, $A_{\rm NS} \equiv 2 \sqrt{R_2}$, 
$B_{\rm NS} \equiv \sqrt{R_2 T_1 T_2} - \sqrt{R_1} (2-T_2)$, 
$C_{\rm NS} \equiv -(\sqrt{T_1} (2-T_2) + \sqrt{R_1 R_2 T_2})$, 
$D_{\rm NS} \equiv (1+R_1)(1+R_2)$, and $E_{\rm NS} \equiv -4 \sqrt{R_1 R_2}$. 
For $L \gg \lambda_F$,  with $\lambda_F$ the Fermi wavelength, and $N T_i \gg 1$ the current is not dominated by a single resonance and the phases $\phi_n$ are uniformly distributed from 0 to $2\pi$. Replacing the sums in Eqs.~(\ref{eq:double1}) and (\ref{eq:double2}) by integrals over $\phi_n$ then yields the average pumped currents
\bea
\langle I_{\rm N} \rangle & = &  I_{\rm C}\, \frac{T_1^{\frac{3}{2}} 
T_{2}^{\frac{3}{2}}}{\pi(1-R_1 R_2)^2} 
\left[ 2\, (-1 + R_1 R_2 + \right. \nn \\ 
& & \sqrt{R_1 R_2 T_1 T_2}) \left. - \pi (\sqrt{R_1 T_2}  + \sqrt{R_2 T_1}) 
\right],
\eea
and 
\bea
\langle I_{\rm NS}\rangle & = & 4\, I_{\rm C}\, \frac{T_1^3 T_2^{\frac{5}{2}}}{\pi (D_{\rm NS}^2 - 
E_{\rm NS}^2)^{\frac{5}{2}}} \left[ 2 C_{\rm NS} D_{\rm NS} 
(D_{\rm NS}^2- E_{\rm NS}^2)^{\frac{1}{2}}
\right. \nn \\
& - & \left.  A_{\rm NS} (E_{\rm NS}^2+2 D_{\rm NS}^2)/2 + 3 B_{\rm NS} D_{\rm NS} 
E_{\rm NS}/2 
\right],
\eea 
whose ratio is plotted in Fig.~\ref{fig:doublebarrier}. As for a nearly-isolated 
quantum dot, $\langle I_{NS} \rangle$/$\langle I_N \rangle$ is largest in case of 
nearly-symmetric coupling. Only for $T_1=1$ the maximum enhancement by 
a factor of $>4$ is obtained, since for $T_1<1$ less electrons reach the NS interface due to normal reflections at the barriers, which in the absence of resonances reduces the effect of Andreev reflection.

\section{Conclusion}

In conclusion, we have studied adiabatic quantum pumping in mesoscopic NS systems. 
Compared to the conductance in these systems we predict two striking differences: (1)
For a nearly-isolated quantum dot with symmetric ($T_1$$=$$T_2$) tunneling barriers,
and a transparent ($T_1$$=$$T_2$$=$$1$) double-barrier NS junction, AR enhances the 
pumped current by a 
factor of 4, which is twice the maximum enhancement of the conductance in these systems. 
(2) In case of quantum pumping this enhancement is not an absolute maximum, whereas 
in case of conductance it is. These differences are due to, resp., the absence of
 an external bias and the asymmetric dependence on the tunnel barriers in case 
of quantum pumping. We hope that these fascinating effects of Andreev reflection on quantum pumping
will find experimental confirmation, e.g. in present-day available nearly-closed quantum dots\cite{folk01}. \newline
Stimulating discussions with C.M. Marcus, N.W.A. Tollenaar and Yu. V. Nazarov 
are gratefully acknowledged. 
This work was supported by the Stichting voor Fundamenteel Onderzoek der Materie (FOM), 
by NFS grant CHE-0073544 and by an NSF MRSEC grant.

\end{document}